# Hybrid quantum nanophotonic devices with color centers in nanodiamonds

SWETAPADMA SAHOO,[1,2,3] VALERY DAVYDOV,[4] VIATCHESLAV AGAFONOV,[5] AND SIMEON I. BOGDANOV[1,2,3*]

[1]*Department of Electrical and Computer Engineering, University of Illinois at Urbana-Champaign, Urbana, Illinois 60801, USA*
[2]*Nick Holonyak, Jr. Micro and Nanotechnology Laboratory, University of Illinois at Urbana-Champaign, Urbana, Illinois 61801, USA*
[3]*Illinois Quantum Information Science and Technology Center, University of Illinois Urbana-Champaign, Urbana, Illinois 61801, USA*
[4]*L.F. Vereshchagin Institute for High Pressure Physics, Russian Academy of Sciences, Troitsk, Moscow, 142190, Russia*
[5]*GREMAN, CNRS, UMR 7347, INSA CVL, Université de Tours, 37200 Tours, France*

*\*bogdanov@illinois.edu*

**Abstract:** Optically active color centers in nanodiamonds offer unique opportunities for generating and manipulating quantum states of light. These mechanically, chemically, and optically robust emitters can be produced in mass quantities, deterministically manipulated, and integrated with a variety of quantum device geometries and photonic material platforms. Nanodiamonds with deeply sub-wavelength sizes coupled to nanophotonic structures feature a giant enhancement of light-matter interaction, promising high bitrates in quantum photonic systems. We review the recent advances in controlled techniques for synthesizing, selecting, and manipulating nanodiamond-based color centers for their integration with quantum nanophotonic devices.

## 1. Introduction

Over the past two decades, diamond has emerged as an outstanding wide-bandgap material for quantum photonic applications [1,2]. Impurity-vacancy (MV) color centers in diamond: NV centers [3] and group IV centers [4] (SiV, GeV, SnV, and PbV), are isolated, optically active atomic defects. MV color centers (CCs) possess a discrete atom-like structure of electronic energy levels, sensitive to local electric, magnetic, temperature, and strain fields. They exhibit bright, photostable single-photon fluorescence in the visible and near-infrared ranges. CCs feature electron spin degrees of freedom that can be coherently coupled to both optical transitions [5] and nearby nuclear spins [6]. Diamond's low free electron concentration, high Debye temperature, and exceptional isotopic purity offer a low-noise environment in the optical and microwave frequency ranges. In this environment, one can harness the unique properties of CCs [4,7] for a plethora of quantum functionalities [8,9], including single-photon sources [10], quantum memories [11–13], quantum gates [14], and sensors [15–18].

The rich functionality of CCs crucially relies on their rates of decoherence and coupling to the electromagnetic fields of interest. The distance from the CC to the diamond surface is an important parameter influencing these rates. For example, deeply embedded bulk CCs ( >100 nm from the surface) exhibit the highest coherence resulting in nearly transform-limited optical linewidth at cryogenic temperatures [19,20] and long spin coherence times, even at room temperature in the case of NV centers [21,22]. However, the weak interactions of electromagnetic fields with deep bulk CCs cause substantial difficulties in performing quantum operations at practical rates [23,24]. The closer the color centers are to the diamond surface, the more engineering options are available to control and enhance their interactions with external electric and magnetic fields, e.g., through optical waveguides and resonators [25–29]. At the same time, such accessibility often comes at the cost of degraded and inhomogeneous color center properties due to the effects of surface and material strain [30]. Nanostructuring bulk diamonds can strongly increase the coupling of optical modes to shallow CCs (within several nanometers from the surface) [31–33]. Shallow CCs can be integrated with other photonic components in all-diamond or hybrid photonic systems. Hybrid integration of diamond into other semiconductor platforms offers reconfigurable architectures, stable optical alignment, a small system footprint, and an interface with the control circuitry [34]. This integration is particularly impactful for established photonic material platforms that lack active quantum functionality.

In nanodiamonds (NDs), CCs can be located within a few nanometers from the diamond surface. ND-CCs can be interfaced with resonant nanoscale optical modes, resulting in a multi-order enhancement of single-photon count rates. In particular, subwavelength ND-CCs are uniquely suitable for coupling with plasmonic resonators, realizing Purcell factors that are several orders of magnitude higher than dielectric resonators [35–37]. The resultant speedup of emission rate up to the THz range will greatly benefit quantum photonic networks and sensors. First, THz emission rates can enable near-THz clock rates and practically viable communication and entanglement rates in quantum networks currently limited to the kHz range and lower [38,39]. Second, faster emission rates may enable the on-demand production of indistinguishable photons at non-cryogenic temperatures [40–42]. Third, THz emission rates could mitigate the effect of frequency mismatches between integrated quantum optical components that often hinder the scalability of quantum networks [42].

Among several types of nanoparticle-based quantum emitters, ND-CCs feature key properties that make them particularly suitable for applications in quantum devices. They can be produced in large quantities, allowing for flexible and rapid prototyping. They are more photostable than colloidal quantum dots (QDs), which typically suffer from photoblinking and photobleaching at high excitation powers [43,44]. ND-CCs are spectrally insensitive to the particle size and feature a relatively small inhomogeneous broadening [45,46]. At the same time, ND-CCs feature much brighter and faster emission than rare-earth ions [47].

The nanoscale size of NDs affects the average optical and spin properties of their CCs. The surface proximity, inhomogeneity in local strain, and material impurities can lead to spectral diffusion [48–50], photoluminescence blinking [52], and degraded spin coherence [52,53]. In addition, the properties of ND-CCs are often inhomogeneous [54]. These properties sometimes conflict with quantum device requirements that may include [10,55–57] photostable emission with near-ideal single-photon purity, narrow fluorescence linewidth with negligible dephasing, near-unity quantum yield, and size compatibility with specific resonator geometries.

Over the past few years, promising results have been achieved for the control of ND-CC properties and their integration into quantum devices. Firstly, the average quality of ND-CCs continues to improve through advanced growth techniques and post-growth surface treatments. For example, CCs in high-purity NDs can feature spin coherence times comparable to shallow color centers in bulk diamonds [58]. Secondly, the selection of NDs with desired optical and structural properties has seen advances driven by microscopy techniques, nanophotonics, and artificial intelligence. Thirdly, device fabrication techniques based on nanoparticle transfer, positioning, and deterministic coupling to nanoscale optical structures have shown promise to realize high-speed quantum devices and sensors integrated on established photonic platforms. This review will survey the recent progress in the synthesis of ND-CCs and the fabrication of ND-based quantum photonic devices (Fig.1). In Section 2, we review the available techniques for the growth (Fig.1 (a)) and post-processing (Fig. 1 (a)) of ND-CCs, with an emphasis on SiVs and GeVs. Section 3 discusses how the heterogeneity in ND-CCs can be tackled through dispersion (Fig. 1(b)), selection (Fig. 1(c)), and manipulation (Fig. 1(d)) techniques. Section 4 discusses ND-based devices featuring enhanced light-matter interactions (Fig. 1(e) and (f)) and their integration into on-chip platforms (Fig. 1(g)).

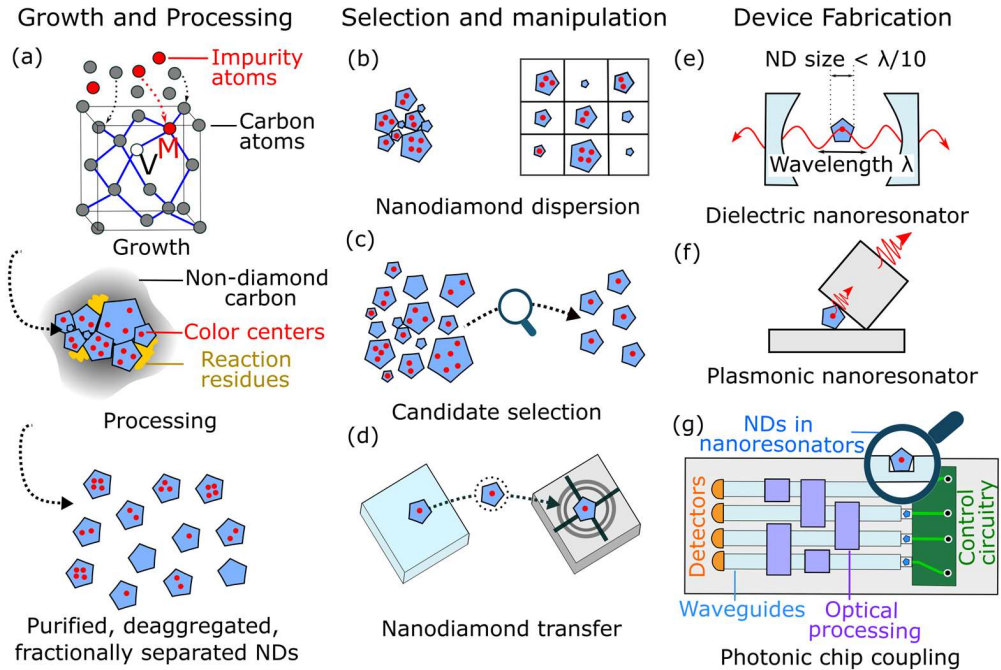

Fig 1. Key procedures for hybrid quantum nanophotonic device fabrication with color centers in nanodiamonds (a) Artificial growth techniques, removal of surface impurities, deaggregation, and fractional separation of NDs containing color centers help control their properties and uniformity. (b-c) Dispersion of NDs and scalable characterization of individual ND-CCs help tackle the heterogeneity in their properties. (d) Controlled delivery of NDs to the desired location enables quantum device assembly. (e-g) Coupling ND-CCs to optical nanostructures enables their interfacing with on-chip photonic circuits.

## 2. Preparation and properties of nanodiamonds with MV color centers

The properties of NDs are determined by specific impurities and structural defects that are strictly individual and store a biographical memory of their formation conditions. The individuality of NDs, in turn, leaves an imprint on the properties of the CCs contained in them. ND-CC properties depend on static inhomogeneities: inclusions, impurities, and structural defects, as well as dynamic thermal and photoinduced effects. The sensitivity of diamond CCs to the local environment leads to a wide statistical distribution in their optical and spin properties, which is exacerbated in NDs by the CC-surface proximity. For quantum photonic device applications [1,56,57], specific ND size, isometric shape, controlled MV type, and concentration is required. Additionally, it is necessary to minimize the presence of electroactive chemical impurities, paramagnetic centers, structural and surface defects, internal strain, and $sp^2$ states of carbon. While MV centers can be found in natural diamonds [59,60], the synthesis and properties of these CCs cannot be controlled. Therefore, the production of ND-CCs meeting the requirements of quantum information applications relies on artificial synthesis methods in atomically controlled conditions. In this section, we review the recent advances in this area, focusing on three sets of tools: methods for the creation of MV centers that are relevant to NDs, fabricating nanoscale diamonds, and approaches for ND post-growth processing. Finally, we discuss how one can draw on these toolsets to obtain ND-CCs with controllable properties.

*2.1 Creation of impurity-vacancy centers in diamond materials*

MV-CCs present two different types of lattice defects in the diamond crystal. The NV center is a complex of an impurity nitrogen atom replacing a carbon atom and vacancy at a neighboring site. This center is characterized by a $C_{3V}$ symmetry and a permanent electric dipole moment [3]. The NV center comes in three charge states: $NV^-$, $NV^0$, and $NV^+$. Negatively charged $NV^-$ centers have been widely used for prototyping quantum devices such as quantum sensors [61,62], and stable solid-state sources of single photons at room temperature [63]. The search for CCs with better optical coherence has turned the spotlight on group IV CCs, including SiV, GeV, SnV, and PbV. Due to significant differences in the sizes of nitrogen atoms and group IV atoms, the structure of group IV CCs is qualitatively different from that of NV centers: the impurity atom is located [4] in the center of a cavity formed by two adjacent vacancies in the diamond lattice, called a "split-vacancy" configuration. Group IV CCs feature a $D_{3d}$ symmetry, characterized by a center of symmetry and the lack of a permanent electric dipole moment [64]. The internal atomic symmetry of these CCs (primarily Si and Ge) greatly suppresses the interaction of orbital states with phonons. It leads to narrow ZPLs at room temperature and the weak dependence of the ZPL frequency on local electric fields.

MV centers can be obtained through the high-temperature annealing of diamond material already containing natural or artificially created impurities and vacancies. Impurity atoms can be introduced into the lattice by ion implantation [65–68] or through the laser-induced Coulomb explosion of a coating layer containing impurities M [69,70]. Irradiation by high-energy electrons, neutrons, or ions introduces vacancies [4,10,56]. Subsequent annealing at temperatures of 700-1100º C leads to the diffusion of vacancies, forming MV centers [56] and relaxing possible structural defects.

The implantation approach allowed for the reliable identification of NV [3,71], SiV [72], GeV [73], SnV [75], and PbV [75] CCs in diamond but leads to structural defects that cannot always be successfully "healed" in the process of high-temperature annealing [76]. This residual damage results in an increased heterogeneity in the optical properties of the CCs.

Alternatively, MV centers can be directly obtained by incorporating impurities during diamond synthesis. The energy of MV defect formation is smaller than those of separate impurities and vacancies, fostering the formation of MV centers [77]. NV, SiV, GeV, and SnV centers can be obtained during detonation, chemical vapor deposition (CVD), and high-pressure high temperature (HPHT) growth [73,78–83]. The concentration of MV defects and the degree of ND crystallinity perfection are determined by the concentration of impurities available during growth, their solubility in diamond at synthesis temperatures, diamond crystallization rates, and the departure from equilibrium during crystallization. Solubility of impurities is determined by the energy of defect formation, which primarily depends on the size of the impurity atom and its binding energy with carbon. As atomic size increases from N to Si, Ge, Sn, and Pb, the M-C bond energy decreases, hampering the incorporation of larger atoms. The atomic size mismatch may be the reason why PbVs have not yet been obtained by impurity incorporation during growth.

*2.2 Obtaining nanodiamond fractions*

Nanoscale diamonds are broadly distinguished into three size categories [78]: (i) Diamondoids - molecules with sizes of 1-2 nm, (ii) single-digit particles with sizes below 10 nm, and (iii) NDs, with a size of 10-100 nm. Diamond particles with a size of 100-1000 nm belong to the submicron range. Due to surface tension, particles smaller than 8-9 nm tend to adopt a spherical shape. A pronounced diamond cuboctahedral faceting appears in nanoparticles larger than 10 nm. Recent reviews discuss various methods for producing nanoscale diamonds [78–80].

The two main ND fabrication routes are fragmentation and condensation. Fragmentation is a top-down route based on the breakup of the input bulk and micro-sized diamonds using mechanical, chemical, and thermal processing. Nano-sized diamond fractions can be obtained through i) high-energy milling of poly- and single-crystal diamonds [78,81], ii) chemical etching of larger (submicron and micro-sized) fractions of diamond, resulting in their gradual shrinking to nanoscale sizes, and iii) graphitization of larger diamond fractions via thermal breakup of surface layers [83] and their subsequent removal by chemical processing.

Condensation is a classic bottom-up route for forming a solid from various phases (gas, liquid, or fluid) of carbon-containing media, resulting from high-energy (thermal, electro-spark, laser) effects. According to the theory of physicochemical evolution of solid matter [85], condensation changes occur in the following sequence: i) the formation of a primary cluster, ii) the formation of a critical nucleus, iii) the growth of nuclei, iv) their aggregation, and v) the structural ordering and maturation of macro-sized particles with a characteristic equilibrium habit. The presence of nanoscale diamonds is limited to the initial stages of the evolutionary route. However, supersaturation of the medium leads to the formation of other metastable nanoscale carbon states (nanographite, fullerenes, spherical and polyhedral onion-like particles, nanoribbons, and various amorphous phases), in addition to NDs. The formation of NDs can occur either via the condensation of carbon atoms and carbon clusters directly into diamond particles, or the phase transition of other metastable nanosized allotropes into diamond [86,87]. Condensation ND synthesis methods include detonation [88], chemical vapor deposition (CVD) technology [89], high static pressures and temperatures synthesis in metal-free growth systems based on different hydrocarbon compounds [90] or carbon nitride [91], laser ablation [92] and ultrasonic cavitation [93].

In addition, NDs can be obtained by direct phase transition of various nanoscale carbon states into diamonds, without atomizing the initial states. Such technologies include methods of ND synthesis based on impact compression of graphite [94], thermal treatment of periodic mesoporous carbon at high static pressures [95], ion irradiation of graphite [97], and electron irradiation of carbon onion-like nanoparticles [97].

Recently, two main HPHT condensation methods based on i) halogenated hydrocarbon growth systems and ii) diamondoid precursors have been developed for ND synthesis [98–100]. ND growth from halogenated hydrocarbon mixtures shows a pronounced dependence of their sizes on the growth mixture processing temperatures. Precise control over particle sizes in the 1-10 nm range has been shown [100]. There are two approaches for ND production using diamondoid precursors. The first approach realizes the idea of chemical overgrowth of molecular diamond-like seeds in a hydrocarbon medium, such as diamondoid molecules [7,101–103]. The second approach is based on ND formation through direct high-pressure polycondensation of pure diamondoids (adamantane, diamantane, triamantane) [99]. These methods currently use diamond anvil cells, limiting large-scale ND production. Because of the similarities in the structure and $sp^3$ hybridization of diamondoids, the conversion into diamond occurs at lower energy and temporal barrier, leading to lower growth temperatures than conventional HPHT methods.

*2.3 Post-processing NDs for controlling properties of MV centers*

Post-synthesis processing techniques can modify the overall ND structure, impurity content, structural defects, surface properties, and internal strain. These techniques are used to purify NDs, control their sizes and improve the color center properties.

NDs are often complex mixtures of diamond, graphitic carbon, or other $sp^2$ carbon allotropes. Two main approaches are adapted to purify and isolate diamond particles: dry processes of plasma treatment or annealing and wet chemical methods of acid treatments. Rapid thermal annealing at temperatures exceeding the standard for $NV^-$ production can efficiently eliminate parasitic, paramagnetic impurities in micron-sized particles [104]. As a result, the degree of $^{13}C$ hyperpolarization improves by an order of magnitude via polarization transfer from optically polarized $NV^-$ centers. Due to the chemical resistance of diamonds, their reaction rates with hydrogen, oxygen, etc. are much slower than those of $sp^2$ allotropes of carbon. Oxidation of NDs can reduce background fluorescence by getting rid of disordered carbon and graphite shells. Air oxidation selectively has been shown to reduce the quenching of CC luminescence. This is evidenced by the elongation of the excited-state lifetime and the increase in photoluminescence brightness (Fig. 2(a)) [105]. Hydrogenation forms C-H bonds at the surface and contributes toward the surface etching of non-diamond carbon and removal of oxygenated groups. Hydrogen plasma treatment leads to the deactivation of CCs near the surface and can improve the photoluminescence stability and increase the proportion of NDs with single CCs [106].

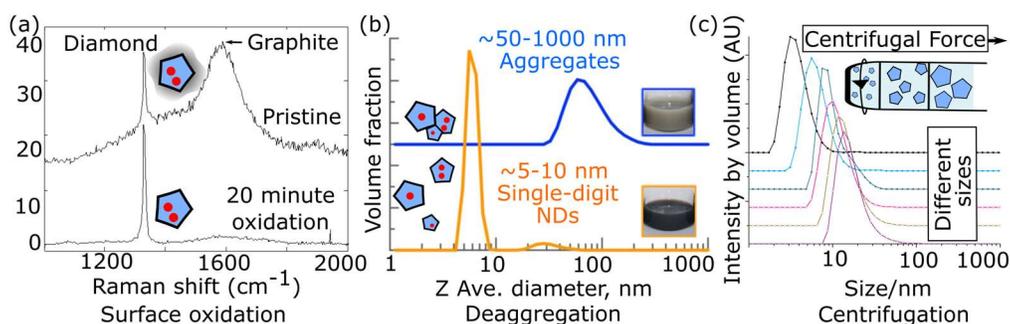

Fig 2. Examples of ND post-growth processing (a) Chemical and oxidation treatment help eliminate non-diamond carbon phases and passivate the ND surface [105]. (b) Salt-assisted ultrasonic deagglomeration breaks up single-digit particle agglomerates in an aqueous solution [107]. (c) Centrifugation separates NDs into different size fractions [108]. Figures reproduced with permission: (a) [105] from © 2010 Elsevier, (b) [107] from © 2016 ACS, (c) [108] from © 2016 ACS.

Three different wet chemical treatment methods are commonly used for purification. The first method uses a 1:3 volume mixture of nitric acid, $HNO_3^+$, with sulfuric acid $H_2SO_4$ or a 1:1:1 volume mixture of $HNO_3^+$/ $H_2SO_4$/$HClO_4$. In this case, purification takes place at a temperature between 100° C and 120° C (the decomposition temperature of $HNO_3$), over a few hours. The second method uses the "Piranha solution", a mixture of sulfuric acid ($H_2SO_4$) and hydrogen peroxide ($H_2O_2$) [109]. The third method uses a mixture of hydrogen peroxide and iron (II) sulfate, called "Fenton" [109]. These three reactions also leave functional groups such as OH, C=O, and COOH on the diamond surface [110]. These surface terminations help in the deaggregation of NDs [111] and improve their dispersibility and stability in solution [112].

Spontaneous agglomeration of NDs, i.e. their tendency to get weakly attached to minimize their surface energy, hampers the isolation of individual particles and obstructs post-synthesis surface modification. Agglomeration has a more pronounced effect in smaller-size fractions [113]. To ensure the NDs are broken up for surface treatments, techniques such as sonication and milling are used [78,114,115]. Effective sonication techniques include salt- (Fig. 2 (b)), [107], sugar- [116] or bead-assisted sonication [113,117]. Surface functionalization of individual NDs with groups that readily dissociate in water increases repulsive forces between them and forms a stable aqueous suspension [118]. Targeted breaking of agglomerates using atomic force microscope probes [119] is useful to free up a particle with a pre-characterized color center from its agglomerate on a substrate.

The wide distribution in the synthesized particle sizes, from sub-microns to nanoscale, is widespread in HPHT systems due to temperature and pressure gradients inside the reaction zone. Fractionation, i.e., size-dependent particle separation, employs two primary techniques. First is membrane filtration, a technique based on the elution and retention of an analyte through membrane pores, typically from 25 to 200 nm. The second is centrifugation, separating particles in a suspension based on a centrifugal force (Fig. 2(c)) that scales with the particle size. The separation of single-digit particles (4 to 6 nm) uses ultracentrifugation [108], with rotation speeds exceeding 15000 rpm [120].

*2.4 Properties of single MV centers in nanodiamonds*

ND-CCs with controllable properties can be obtained through combinations of the previously described techniques for ND synthesis, color center production, and post-growth treatment. Currently, there are three main approaches for obtaining ND-CCs. In the first approach, CCs are created initially in macro or micro-sized diamond materials that undergo subsequent fragmentation [121–123]. The second approach involves creating CCs in synthesized NDs [124,125]. The third approach is the directed synthesis of individual NDs with the inclusion of MV centers during growth, most commonly performed through i) HPHT "homogeneous" diamond nucleation or overgrowth of diamond seeds (Fig. 3(a)) [98,126,127], ii) CVD overgrowth of diamond seeds (Fig. 3(b)) [129], and iii) detonation (Fig. 3(c)) [130]. Additionally, ND-CC synthesis via HPHT overgrowth of diamondoid molecules containing dopant atoms in a hydrocarbon medium has been realized [101]. This method is particularly promising for obtaining fluorescent NDs with highly uniform dimensional and optical parameters and a controllable number of specific optically active CCs per particle.

In the rest of this section, we give a few examples of ND-CC properties obtained using the techniques described above.

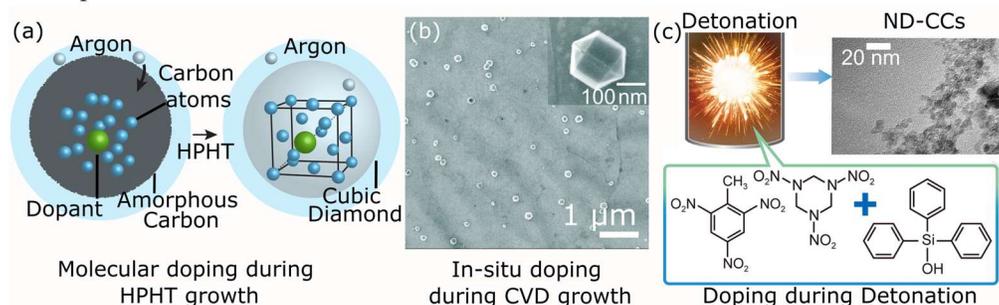

Fig 3. Approaches for the artificial synthesis of ND-CCs. (a) Synthesis of a doped amorphous carbon precursor and transformation at high temperatures and pressures results in color centers homogeneously distributed within the ND [127]. (b) Spatially isolated NDs with high crystalline quality can be obtained via microwave-plasma-assisted chemical vapor deposition (MPCVD) [128]. (c) Detonation of organic matter produces nanodiamonds that are uniform, small and spherical [130]. Figures reproduced with permission: (a) [127] from © 2019 AAAS, (b) [128] from © 2011 IOP Publishing, (c) [130] from © 2021 Elsevier.

In the first demonstration of ND-based single photon sources with NV centers [131], NDs (~40 nm) were obtained from synthetic diamonds subjected to electron irradiation and subsequent high-temperature annealing. Single NVs featured an excited state lifetime of ~25 ns, significantly longer than that in bulk diamonds (11.6 ns) [132]. Single-digit NDs in the size range 5–7 nm obtained by detonation [110], electron irradiation, annealing [135], and subsequent fragmentation of microsized diamonds also yielded single NV centers. However, the optical transition frequency was inhomogeneous due to the increased local strain interacting with the NV center's DC electric dipole [136]. A broad emission spectrum extending from 637 nm, corresponding to the ZPL of the fluorescence line, to 800 nm is seen. A significant discrepancy exists between the ZPL linewidth and spectral diffusion values for NV-NDs, where they are ~ 1.2 GHz and 16 MHz, respectively [49,134]. Studying temporal photoluminescence intensity fluctuations of ND-CCs can reveal the dynamics of surface and impurity-induced photoblinking and photobleaching. The effect of the surrounding environment can be harnessed to inhibit the photoblinking of ND-CCs [135]. It has been shown that embedding 5-nm sized NDs in a polymer film can mitigate NV center intermittency [51].

**Table 2. The reported properties of sizes, ZPL wavelength and linewidth, fluorescence lifetime, and saturation photocounts for SiV and GeV doped nanodiamonds**

| Color centers | Fabrication method | ND sizes | ZPL center (nm) | Linewidth (nm) | Fluorescence lifetime (ns) | $I_{sat}$ (kcps) | Ref. |
|---|---|---|---|---|---|---|---|
| SiV | SiV-ND-MPCVD-Ir | 90-170 | 737.6-740.9 | <2.2 @ RT | 0.2-2 | 260-4800 | 165 |
| SiV | SiV-ND-MPCVD-Si | 100 - 300 | 737.3 – 739.0 | 210 - 850 MHz @ 4K | 0.8 – 2.3 | 9 – 92 | 155 |
| SiV | SiV-ND-HPHT-CH | 10 - 30 | 727 - 747 | 2.1 – 2.9 @ RT | 0.9 – 3.8 | 18 – 200 | 166 |
| SiV | SiV-ND-HPHT-CH | 5 - 15 | 737.8 – 738.3 | 5.1 – 6.7 @ RT | | 19 –143 | 167-168 |
| GeV | GeV-ND-HPHT-CH | 10 - 50 | 602.4 – 604.6 | 5.2 ± 1.2 @ RT | 10 - 20 | | 173 |

The first systematic studies of NDs with single SiV centers were carried out using microwave-plasma-assisted CVD growth using synthetic NDs as seeds on iridium substrate (referred to SiV-ND-MPCVD-Ir) [136]. The studied samples had well-faceted diamond single crystals with an average size of 130 ± 40 nm. The SiVs featured bright photoluminescence with ZPL saturation counts up to 4.8 Mcps, exhibiting a clear fine structure of the photoluminescence spectra in single SiVs.

Later, single SiV centers were obtained by growing diamond nanoseeds with sizes of 4-6 nm on a silicon substrate by the MPCVD method (referred to SiV-ND-MPCVD-Si) [137]. The high-resolution photoluminescence excitation technique allowed selective excitation of individual CCs from ensembles of SiV centers, showing an excited-state lifetime of 0.82-2.29 ns at 4K. High fluorescence stability and low spectral diffusion were combined with a narrow ZPL linewidth of 325 ± 30 MHz, about 2.3 times the lifetime-limited linewidth of 141 MHz.

NDs with single SiV centers have also been obtained based on halogenated hydrocarbon HPHT growth systems (referred to as SiV-ND-HPHT-CH(F)) [138–140]. Naphthalene ($C_{10}H_8$) and tetrakis(trimethylsilyl)silane ($C_{12}H_{36}Si_5$) were used as the hydrocarbon and silicon-containing components of the initial growth mixtures. Fluorinated graphite ($CF_{1.1}$) and octafluoronaphtalene ($C_{10}F_8$) were used as additional components of the mixture, contributing to a decrease in the temperature threshold of diamond formation and an increase in the ND yield [87,98]. The Si/C atomic ratio has been shown as a viable tool to control the proportion of NDs with single SiV centers. This approach allowed to tune that proportion from 2.5% to 30% in the 10-30 nm and 50-100 nm fractions of SiV-ND-HPHT-CH(F)[138]. The width (FWHM) of most ZPL peaks of single SiV centers was 2.1–2.9 nm. Lower widths indirectly indicate higher degrees of structural perfection and homogeneity of CC properties than that of ND-CCs with higher ZPL widths. The ZPL frequency and fine structure of ND-CCs are primarily determined by the magnitude and direction of the internal strain [141,142]. Strain along the MV center axis does not break the symmetry of the center. As a result, only the ZPL energy is changed, while the ground and excited level splitting remain unaffected. In contrast, a transverse strain breaks the symmetry and splits both the ground and excited levels of the MV center. An analytical model based on this consideration has highlighted that most single centers in SiV-ND-HPHT-CH(F) are characterized by low degrees of internal strain, comparable to that in the best of unstrained bulk diamonds samples [106]. This hypothesis is corroborated by ZPL fluorescence spectrum linewidths close to the lifetime limit [45]. However, a high average lattice quality of SiV-ND-HPHT-CH(F) type nanodiamonds does not preclude noticeable heterogeneities in the opto-physical properties of individual SiV-NDs. The smallest fractions of SiV-ND-HPHT-CH(F) featured a size distribution of 7.5 ± 5 nm, as measured with TEM [139,140]. The narrow ZPL frequency distribution (738.06 ±0.27 nm), X-ray diffraction, and high-resolution TEM data point to a high degree of crystallinity. The influence of NDs treatment with hydrogen plasma on the optical properties of SiV-CCs is investigated in the work [106]. A study of the fluorescence characteristics of NDs in the mode of non-resonant excitation at 532 nm at a temperature of 8 K showed that surface treatment with hydrogen plasma leads to a decrease in the level of spectral diffusion, an increase in fluorescence stability, and allowed spectrally resolving the fine structure of the SiV centers. This type of treatment leads to the deactivation of CCs located near the surface, reducing the total number of active centers but increasing the relative number of ND with single SiV centers.

Thermal transformations can result in nanodiamonds with single GeV centers at high pressures in heteroorganic growth systems with doping elements [143,144]. Single GeVs were obtained through HPHT synthesis of NDs (~50 nm) using a binary mixture of adamantane ($C_{10}H_{16}$) and tetraphenylgermane ($C_{24}H_{20}Ge$) [143]. Another growth method, referred to as GeV-ND-HPHT-CH, based on homogeneous mixtures of naphthalene ($C_{10}H_8$) with tetraphenylgermane ($C_{24}H_{20}Ge$) yielded 10-50 nm NDs with single GeVs, whose optical characterization is detailed in [144]. The ZPL linewidth value of $5.2 \pm 1.18$ nm is comparable to the GeV center in bulk diamond at room temperature [145]. The narrow distribution of the ZPL line of the photoluminescence spectrum of GeV centers ($603.5 \pm 1.1$ nm) indicates weak spectral diffusion and low internal stresses. Maximum saturation counts varied from 0.2 to 1.5 Mcps- about 10 times higher than those observed in bulk diamond samples [73].

## 3. Isolation and characterization of nanodiamonds

Once NDs are grown, ensemble measurements can be performed directly on a colloidal solution to determine the state of deaggregation [107], average particle size [148], or ensemble CC fluorescence spectra [147], among other properties. However, a wealth of useful information hidden in average measurements can be revealed by characterizing large sets (hundreds or more) of individual NDs. Large-scale individual ND characterization provides detailed feedback on growth techniques and post-growth treatment [147,148]. Individual ND screening is also indispensable to select candidates with desired properties from a wide pool of heterogeneous emitters for the fabrication of quantum photonic devices. In this section, we first review the techniques enabling the isolation of NDs on a substrate, focusing on the realization of regular arrays. Then, we discuss techniques compatible with a rapid, large-scale characterization of individual ND-CCs.

### 3.1 Isolation of individual nanodiamonds

Before assessing the properties of individual NDs and color centers, the particles are typically dispersed and isolated on a flat substrate. NDs can be spread using drop-casting, i.e. depositing a solution of NDs on a substrate and drying it under controlled pressure and temperature [149] (Fig. 4(a)). Alternatively, spin-coating may be used to eliminate the solvent by centrifugal force. In both methods, the positioning of NDs is random with a spatially inhomogeneous concentration [150]. Instead, regular arrays can be achieved by controlling the ND-substrate interactions [151,152]. In such arrays, NDs occupy pre-determined microscope-resolvable sites, with a recordable position, facilitating the identification of the same particle over various stages of work. Such arrays allow large-scale and automated particle characterization, identification, and transfer. We will now review three approaches to forming such arrays depending on the mechanism of ND attachment to the substrate: (i) chemical bonding, (ii) capillary forces, and (iii) electrostatic attraction.

Chemical functionalization of ND and substrate surface promotes ND-substrate attachments via strong covalent or ionic bonds. Furthermore, if the substrate functionalization layer is patterned, then regular ND arrays can be formed [153]. For example, such arrays have been demonstrated based on covalent bonding between carboxylic (-COOH) ND terminations and amine-terminated substrate patches (Fig. 4(b)) [154]. This method is largely insensitive to the substrate material and surface geometry. The yield of this strongly bonded self-assembly mechanism can exceed 90% by controlling the ND concentration [154]. Robust chemical bonding may be preferred when subsequent sample cleaning or processing is intended.

Arrays can also be created by depositing a solution of NDs onto an ordered hole arrangement lithographically defined in a sacrificial layer on a substrate. NDs are driven into the holes by capillary forces and remain at their locations after the lift-off of the sacrificial layer (Fig. 4(c)) [155,156]. Lithographic template-assisted techniques allow array assembly of as-is NDs, regardless of their surface terminations. The ND-substrate interaction, in this case, is mainly due to van der Waals forces. The weak bonding of NDs is advantageous for their subsequent manipulation, e.g. the transfer to other substrates [157]. This method features a median yield of individually occupied trap sites of 76% [156].

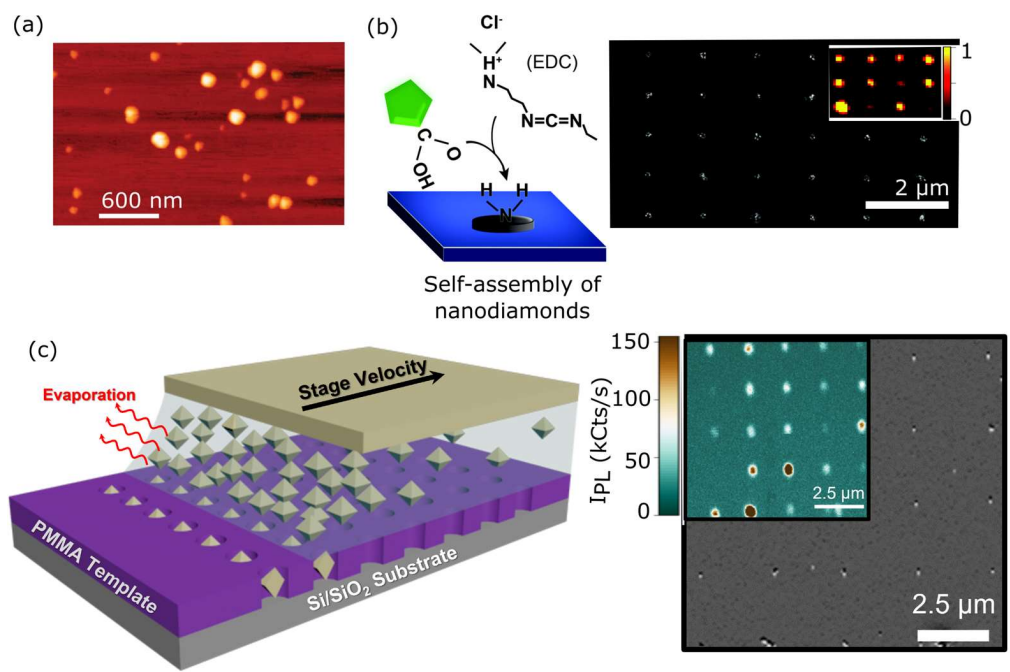

Fig 4. ND dispersion techniques (a) AFM scan showing random dispersal by drop-casting a diluted ND solution on a substrate [149]. (b) Schematic of ND array assembly assisted by chemical functionalization (left). SEM image (right) and the corresponding PL map (right inset) of a patterned area [154]. (c) Template assembly of NDs into lithographically defined apertures (left). AFM image and PL scan of a 5x5 ND array (right) [156]. Figures reproduced with permission: (a) [149] from © 2009 ACS, (b) [154] from © 2016 RSC, (c) [156] from © 2022 ACS.

Treating NDs and the substrate surface with oppositely charged terminations promotes attachment via Coulomb interactions. In this situation, the electrostatic force can be leveraged to drive NDs towards the substrate [158,159]. The substrate can be patterned into arrays of electrostatically charged sites to control the ND attachment positions. A polarized AFM tip can be used for patterning on an electret spin-coated on a conductive substrate [160]. By controlling the surface potential of patterned charge dots, one can achieve the attachment of a single ND per dot. Alternatively, using the electrohydrodynamic (EHD) dispensing method, one can position nanodroplets containing a programmed quantity of charged NDs on an oppositely charged substrate. Once an ND-laden nanodroplet lands on the substrate, the solvent rapidly evaporates, leaving ND clusters behind, with up to 98% yield [158].

Electrostatic assembly and lithography can be combined to precisely position individual NDs onto electrostatically charged pad-like structures. The surface charge of the pads can be manipulated by dipping the substrate into a solution. If the pH value of the solution is above the isoelectric point (IEP) of the pads, they accumulate a positive surface charge and vice-versa. By ensuring that the substrate accumulates the opposite charge from that of the pads, the combined effect of pad attraction and substrate repulsion allows for selective and nanoscale precise positioning of NDs. In [159], 48% of the pads had a single ND. NDs can be positioned on the same array multiple times by cleaning and redepositing.

*3.2 Characterization of nanodiamond properties*

The ND parameters to be characterized largely depend on the experimental needs. In particular, rapid, scalable characterization techniques are necessary for large sets of NDs (hundreds or more) and when several time-consuming measurements have to be performed on each particle. We now discuss how such characterization can be enabled by (i) speeding up the data acquisition and (ii) optimally using the available data through various instances of machine learning.

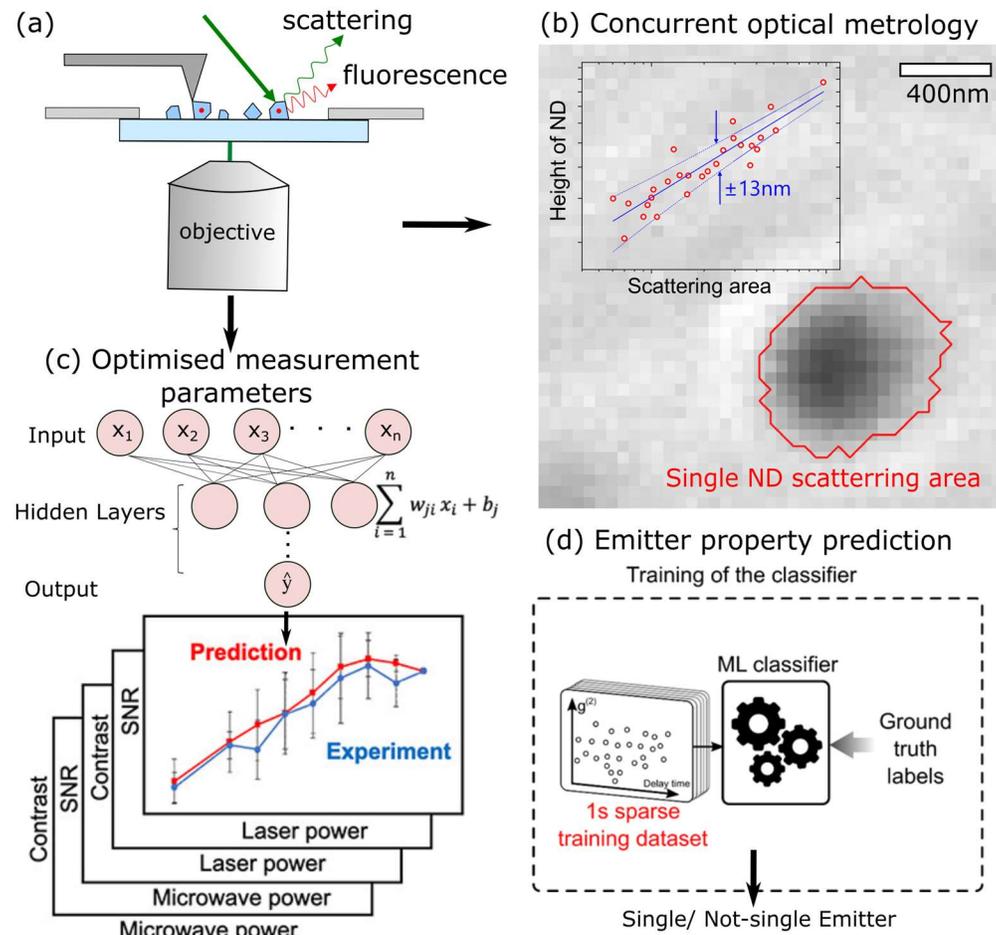

Fig. 5. Large-scale, high-throughput characterization of ND-CCs is useful for statistical studies of growth and post-growth treatments and device fabrication. (a) Schematic of a typical characterization setup. (b) Sizing throughput can be maximized by replacing slow traditional individual ND sizing methods with all-optical metrology [161]. Supervised machine learning accelerates time-costly experiments by optimally using data for (c) predicting optimal measurement parameters (e.g. optimal excitation power for spin characterization [162]), and (d) measurement of ND-CC characteristics (e.g. single-photon purity [163]). Figures

reproduced with permission: (b) [161] from © 2021 AIP, (c) [162] from © 2021 ACS, (d) [163] from © 2020 Wiley-VCH.

The integration of NDs into nanoscale photonic structures is highly sensitive to particle size. For example, the resonant frequency and the photonic density of states of gap plasmonic resonators such as nanopatch antennas strongly depend on the gap width (typically below 100 nm) that is ultimately limited by the size of the object to be placed in the gap [164]. ND height and lateral size can be measured by atomic force microscopy (AFM) and tunneling or scanning electron microscopy (TEM/SEM) respectively with a typical precision of less than 1 nm. However, slow AFM scanning speed and beam-induced sample contamination in electron microscopy methods [165] preclude the use of these methods for scalable characterization. Despite the deeply subwavelength size of NDs, optical techniques have shown promise for their sizing. Furthermore, optical sizing can be concurrently used with the quantum optical characterization of the emitters. For example, linear pump scattering allows the sizing of NDs down to 35 nm with a precision of ±13 nm, when provided with a preliminary absolute calibration by AFM (Fig.5(b)) [161]. Alternatively, coherent anti-Stokes Raman scattering (CARS) allows the sizing of individual NDs down to about 30 nm [166].

Machine learning (ML) approaches are suitable for optimizing multivariate ND-CC characterization variables that have a non-linear effect on measurements [167]. For example, ML algorithms such as linear regression, neural networks, and random forests have been used to predict the optimal laser power and the microwave powers to obtain the highest signal-to-noise ratios and contrast in optically detected magnetic resonance experiments of NV-NDs (Fig.5(c)) [162]. The use of ML can also speed up time-costly measurements through classification and regression of photophysical emitter parameters from relatively sparse data. Supervised ML algorithms have been used to rapidly classify ND-CCs according to their value of $g^{(2)}(0)$ with respect to a specified threshold [163], with an acquisition time of 1s and an accuracy of over 90% (Fig.5(d)). Furthermore, regressive convolutional neural networks have been used to estimate the value of $g^{(2)}(0)$ from sparse data, reaching a precision of less than 0.05 over an acquisition time of 5 s [168].

## 4. Nanodiamond manipulation and transfer

Heterogeneous integration of diamond color centers promises the realization of quantum information systems on robust photonic platforms [169]. Due to the nanoscale size of NDs, their interfacing with optical resonators and on-chip waveguides requires the development of specialized tools. In this section, we review tools that allow the transfer and movement of NDs for their heterogeneous integration into photonic devices.

Probe-assisted manipulation techniques using AFM [157,170] or nanomanipulators integrated into an SEM chamber [171] (Fig. 6(a)) have become increasingly popular for these tasks. AFM offers precise nanomanipulation through (i) deterministic pick-and-place transfer of NDs, and (ii) pushing NDs along the substrate surface [Fig. 6(b)]. During AFM-assisted transfer, one applies a downward force with the probe to attach the ND to the probe tip. The probe then transfers the attached ND to the target position and releases it by pressing it against the substrate. The particle-tip adhesion can be controlled by choosing tip radius using robust semi-adhesive tip coatings [172]. For transparent substrates, the optical fluorescence signal can be used as feedback in both the source and target areas, provided that the AFM is interfaced with a confocal optical microscope and both systems are independently controlled and coordinated [157]. In opaque conductive substrates, long-range Coulomb interactions allow the mapping of the target area in the non-contact mode and can guide the ND placement [173]. After transfer, the ND position can be adjusted by pushing it with an AFM tip in contact mode [170,174]. Intersubstrate transfer of NDs is also possible using nano-manipulators, integrated into an SEM chamber, offering nanoscale precision and real-time monitoring [171]. Pick-and-place transfer of NDs from the substrate to nanofibre-coupled systems has been demonstrated using glass probe manipulators with real-time optical imaging, albeit with microscale precision [175,176].

The control of optical and electric forces offers ND nano- and micro-manipulation and sorting in liquids and vacuum. Optical tweezers (OTs) [177] can trap and manipulate NDs in solution using a focused laser beam, providing positional control for coupling. Optical forces can be used to select and sort ND-CCs according to their properties. By exploiting the balance of resonant absorption and scattering forces induced by counterpropagating detuned lasers in a single-mode fiber, selective transportation of NDs containing resonant NV centers and non-resonant NDs with almost no NV centers in an aqueous solution has been achieved (Fig. 6(c)) [178]. Additionally, laser power-dependent optical transport of NDs enables size-sorting inside microfluidic channels [179]. Recently, electrothermoplasmonic forces have been used to place NDs into the hot spots of plasmonic nanoantenna modes [180] (Fig. 6(d)). In this approach, an AC electric field combined with laser illumination creates fluid vortices carrying particles at speeds on the order of 10 μm/s. In seconds, NDs are delivered to and trapped in plasmonic hot spots that act similarly to the optical tweezers, but with a higher trapping potential due to strong electric field enhancement [181,182]. The controlled orbital motion of single NDs with a trajectory radius of 28 nm, has been demonstrated by angular momentum transfer from circularly polarized incident light via a gold trimer nanoantenna [183].

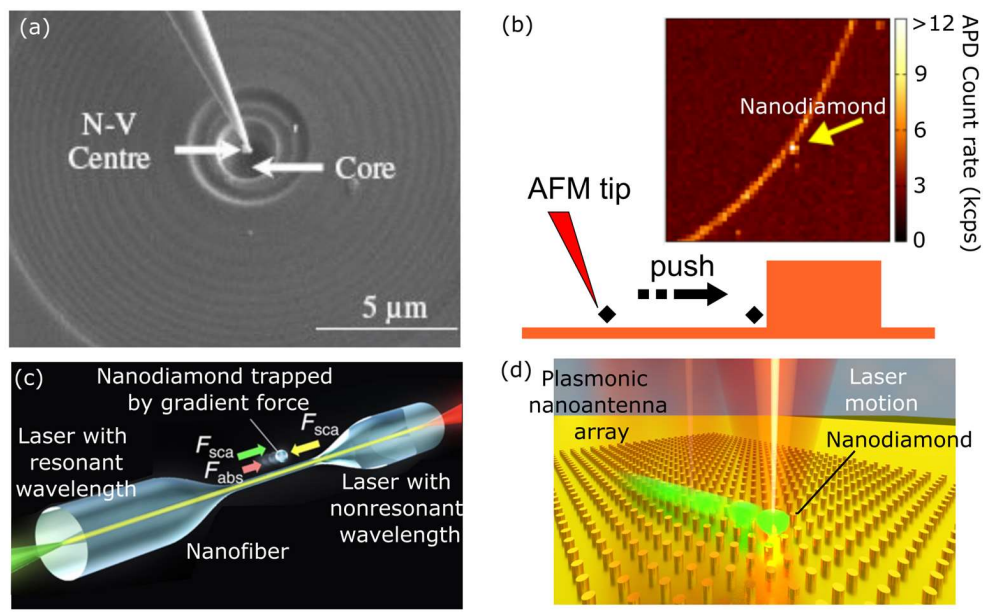

Fig. 6. ND transfer and manipulation for device fabrication. (a) AFM-assisted transfer of pre-characterized NDs to a target substrate [184]. (b) AFM-assisted coupling of an ND to a microring resonator [174]. (c) Optical selection and sorting of NDs using a nanofiber immersed in water [178]. (d) Using electrothermoplasmonic flow, a single ND can be transported toward a plasmonic hot spot within a few seconds [180]. Figures reproduced with permission: (a) [184] from © 2009 OSA, (b) [174] from © 2019 IEEE, (c) [178] from © 2021 AAAS, (d) [180] from © 2021 ACS.

## 5. Integration of color centers in nanodiamonds with nanophotonic structures

The integration of quantum emitters with compact, reconfigurable, and stable photonic circuits promises functional high-bitrate quantum optical systems [185]. Such symbiotic integration equips most photonic platforms with the missing active quantum functionality and, conversely, improves the optical properties of quantum emitters. Optically, mechanically, and chemically robust ND-CCs are compatible with any material platform. Their nanoscale sizes allow high coupling rates, up to the THz range. In this section, we review the integration of ND-CCs with dielectric and plasmonic nanoresonators and on-chip waveguides.

The intrinsic excited-state radiative lifetime in bulk diamond CCs is on the order of 10 ns [4,132], which is compatible with single-photon clock rates of < 10 MHz. The absence of high-index surrounding bulk material in subwavelength NDs slows the vacuum excited-state radiative lifetime by about 16 times [186]. The long lifetime further restricts the achievable single-photon clock rate and leads to an order of magnitude disparity in quantum yield [187,188]. Efficient collection of quasi-isotropic emission from ND-CCs requires bulky confocal microscopy setups.

## 5.1 Integration with nanoresonators

Nanoscale optical resonators can strongly enhance the interaction between optical dipoles of ND-CCs and photonic modes carrying control and fluorescence signals. The Purcell factor quantifies this enhancement in the weak interaction regime [189]. By improving photon emission rates [190], the Purcell effect in nanophotonic resonators provides several additional benefits. If the decay rate is larger than the dephasing rate, then the temporal photon indistinguishability improves [191,192]. Because it selectively enhances the radiative emission rate over the nonradiative rate, the Purcell effect improves the emitter's quantum yield [193,194]. Additionally, interfacing nanoresonators with specific optical modes, e.g. through a photonic waveguide, can improve photon collection efficiency [195,196]. Waveguides help connect the CCs to the rest of the circuit by facilitating their interfacing with on-chip controls: optical pumps, RF, and DC fields [197], and routing the collected fluorescence to other devices and chips [198].

High Purcell factors can be achieved by optimizing the trade-off between the resonator's quality factor, $Q$, and the mode volume, $V$, under the boundary conditions imposed by the properties of emitters and materials. In dielectric resonators, $V$ is bound by the diffraction limit on the order of $V_{diff} = \left(\frac{\lambda}{2n}\right)^3$. For diffraction-limited mode volumes, $Q$ can be as large as $10^3$-$10^4$ but may be in practice limited by the emitter's linewidth [199]. ND-CCs can be integrated with dielectric resonators of various geometries and material compositions, including microdisks, microspheres, and photonic crystal cavities (PCC) made from insulators and wide-gap semiconductors. These approaches show particular promise at cryogenic temperatures, where the CC linewidths spectrally fit into high-Q dielectric resonances [200]. The photoluminescence signal of ND-CCs has shown an increase of 2-5 times when cryogenically coupled with whispering gallery modes in SiC [198]. High-Q resonators ($Q > 10^8$) made from silica with low autofluorescence made it possible to observe strong coupling dynamics with single ND-NVs [201]. ND-NVs integrated with GaP PCCs (Fig. 7(a)) have featured a Purcell factor of ~12 at non-cryogenic temperatures [202].

Plasmonic resonators rely on the deeply subwavelength confinement of light rather than on high-quality factors to offer broadband, strongly enhanced light-matter interactions [203]. Being subject to scaling laws different from those in dielectric resonators, plasmonic resonators have orders of magnitude higher fundamental limit on the achievable Purcell factor [37]: while $Q$ usually does not exceed $10^1$-$10^2$, the mode volume can be $10^4$-$10^5$ times smaller than the diffraction limit. In particular, attaining spontaneous emission rates in the THz range [193] may allow the production of indistinguishable photons even at non-cryogenic temperatures [37,42]. Modest Q-factors and, consequently, broader bandwidths of plasmonic resonators are compatible with those of emitters, up to room temperatures [204,205].

The major shortcoming of plasmonic nanostructures is the high optical loss in metals [206] resulting from i) direct dissipation of dipole energy to heat, at short dipole-metal distances, called quenching [207], and ii) absorption of emitted plasmons. Nanostructures with gap surface plasmon modes alleviate the quenching problem as in these structures, plasmon emission rates scale together with the quenching rates, even at single nm dipole-metal distances [208]. Plasmon absorption losses can be minimized by using epitaxial metal nanostructures and films [209] and by designing impedance-matched plasmonic nanoantennas that rapidly radiate the plasmon energy, ideally overcoming plasmon absorption rates [210].

A nanopatch antenna (NPA), consisting of a metallic nanocube and metallic film, combines the idea of a gap plasmonic resonator with that of an optical antenna [211]. Large enhancements of quantum emission have been observed at room temperature in colloidal quantum dots [212] and ND-CCs [36] placed in the gap of gold and silver NPAs (Fig. 7(b)) made by random assembly [36]. The use of crystalline silver in NV-coupled NPAs leads to radiative efficiencies of about 20% and an average detected saturated intensity of up to 35 million photon counts per second. Furthermore, crystalline silver NPAs have been deterministically assembled around single pre-selected NDs, achieving optimal relative ND-nanocube positioning [184]. The ND volume limits that of the gap resonator in NPAs, but through controllable photomodification of the nanocubes, the gap can be reduced, further enhancing the radiative emission rate [193]. Unlike NPAs, dimer plasmonic nanoantennas can be dipole-aligned with single ND-CCs during assembly to achieve optimal CC-to-antenna coupling and linearly polarized emission [35].

*5.2 Integration with fibers and waveguides*

Nanoresonators are comparable to or smaller than the optical wavelength and, therefore, inherently suffer from poor emission directionality, especially in the case of plasmonic resonators. To interface resonator-enhanced ND-CCs with other devices, coupling to structures with high directionality is needed. Depending on the desired outcome, we distinguish the problems of coupling the emitted radiation into a fiber and that of coupling into an on-chip waveguide.

ND-CC coupled single-mode optical fibers [213] and photonic crystal fibers [214] are promising for fiber-network integration. Incorporating plasmonic bullseye enhanced ND-CCs promises directional emission into fibers. The fluorescence needs to be directed out of the substrate plane. The numerical aperture (NA) is on the order of 0.1 for single-mode fibers and 0.3-0.5 for multimode fibers at the emission wavelengths of ND-CCs. Coupling NV-NDs to plasmonic bullseye antenna promises over 50% coupling efficiency into a multimode fiber [215]. Similar work was carried out on SiV-NDs on a metallic bullseye design, where it was possible to saturate optical transitions with a low NA objective of 0.25 [Fig. 7(c)] [216]. In a simulation, combining plasmonic resonators with bullseye antennas [217,219] leads to high directionalities of about ~85% for a 0.9 NA objective.

Coupling the emission of ND-CCs with waveguides is necessary to realize the efficient transfer of photons and provide an interface to other photonic elements [174]. Deterministic integration of NV-NDs with free-standing dielectric rib waveguides, with a coupling efficiency of ~0.05, has been shown on a silica-on-silicon platform [217]. Waveguide embedded nanoresonators with ND-CCs have been explored for realizing on-chip photonic integrated circuits [218,219]. Single $NV^-$ and $SiV^-$ centers coupled with high-Q $Si_3N_4$ photonic crystal cavities (PCCs) in a crossed-waveguide design [220,221] have achieved up to 71% coupling efficiency for NVs and a Purcell enhancement of >4 for SiVs, at 4K. Coupling ND-CCs via integrated nanophotonic optical cavities can enhance the emission into ZPL and funnel the emission into the guided optical modes of the waveguide. NV-NDs have been coupled with photonic platforms such as $(Ta_2O_5)$-on-insulator on a silicon chip [Fig. 7(d)], with grating couplers, low-loss waveguides, and a 1D PhC-cavity providing a Purcell factor of ~3.3. A coupling efficiency of 77% is shown, allowing the collection of about $3.3 \times 10^5$ photons at room temperatures [222].

A quantum plasmonic launcher (Fig. 7(e)) realizes the first step towards the integration of ND-coupled plasmonic nanoresonators by featuring predominantly in-plane emission [223]. The structure consists of ND-CCs sandwiched between two silver films. Launchers containing single NV centers have shown an average fluorescence lifetime shortening of about 7000 times, with more than half of the emission coupling into in-plane surface plasmon polaritons (SPPs). The integration of NV-NDs with gap-plasmon waveguides has shown a $\beta$ factor of 0.82 and decay rate enhancement of ~ 50 [224]. Propagation lengths as high as ~33 $\mu m$ have been shown in GeV-NDs integrated on-chip with dielectric-loaded plasmonic waveguides (Fig. 7(f)) [225]. For a more detailed review of quantum-emitter plasmonic waveguide coupled structures, the reader is referred to Ref. [226].

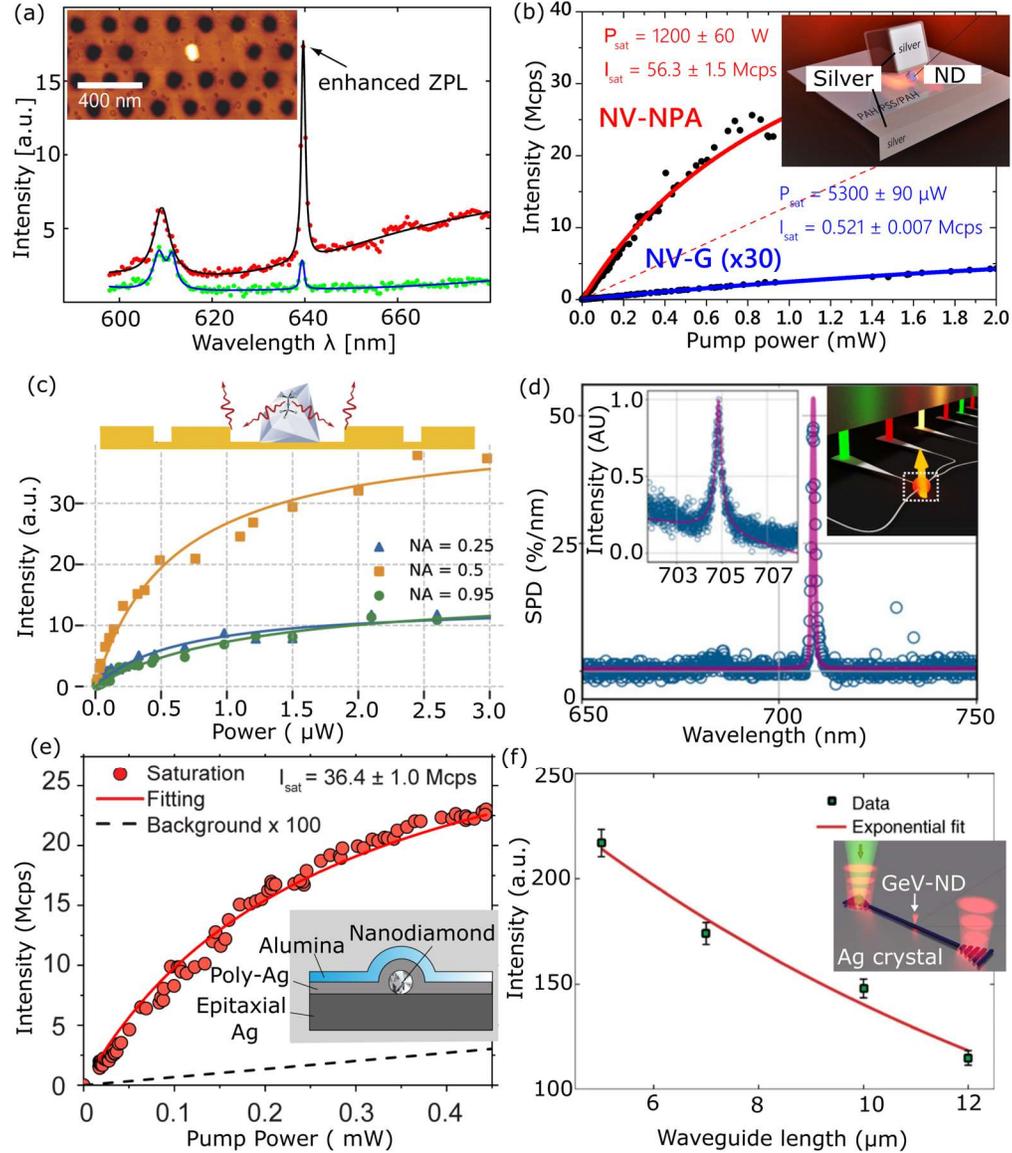

Fig 7. Integration of ND-CCs with various nanophotonic structures. (a) ZPL enhancement (red dots) when ND is coupled to a GaP cavity (inset) as compared to cavity without ND (green dots) [202]. (b) Fluorescence saturation curves showing an increase of two orders of magnitude in saturation counts for an NV-ND coupled to a nanopatch antenna (inset), compared to an NV-ND on a coverslip glass substrate [36]. (c) Saturation measurements of SiV-ND integrated with a metallic bullseye antenna (top) under resonant drive is possible with low-NA optics of 0.25 [216]. (d) Photoluminescence of an integrated NV-ND with $Ta_2O_5$ waveguides (inset) after transmission through a grating coupler is dominated by a single resonance [222]. (e) Saturation

measurements of an NV-ND sandwiched between two silver films forming a quantum plasmonic launcher (inset), which routes a significant fraction of the emission into in-plane SPPs [223]. (f) A GeV-ND integrated with dielectric-loaded plasmonic waveguides on crystalline silver flakes (inset) features a propagation length of 11.8 $\mu$m [225]. Figures reproduced with permission: (a) [202] from © 2010 AIP, (b) [36] from © 2018 ACS, (c) [216] from © 2021 IOP Publishing, (d) [222] from © 2020 ACS, (e) [223] from © 2020 Wiley-VCH, (f) [225] from © 2018 Nature Portfolio.

## 6. Discussion

The combination of controlled synthesis, rapid characterization, and high-yield manipulation of ND-CCs constitute a promising path to scale up the fabrication of ND-based quantum photonic devices and realize high-bitrate quantum networks at non-cryogenic temperatures. The synthesis of high-quality, homogenous NDs is a critical step toward scalable integrated devices based on ND-CC. A rapid large-scale search for ND-CCs with desirable quantum properties would benefit from the future development of multi-parameter measurements [227], ML-assisted optimal processing of sparse data, and parallel characterization of multiple particles [228].

For large-scale ND-CC integration, rapid manipulation techniques with high yield and throughput are the need of the hour [229]. Automated manipulation with precise positional and orientation control, simultaneous multi-particle manipulation, and resource-optimal imaging [230] can address these needs. ND-CC integration yield will benefit from approaches that grant spatial misalignment tolerance without compromising mode volume [231].

In the near future, ND-CCs may realize room temperature strong coupling [232] or plasmon-assisted generation of indistinguishable photons [233], further extending their quantum optical functionality. External optical control to mitigate spectral broadening [234] can ensure indistinguishable photons from heterogeneous ND-CCs. Inverse design [235] and machine learning techniques promise rapid automated design of ND-CC integrated devices with higher performances and smaller footprints.

**Acknowledgments.** SS and SIB acknowledge funding from the Strategic Research Initiative (SRI) program of the Grainger College of Engineering at the University of Illinois at Urbana-Champaign. The authors thank Joshua Akin for his help with the preparation of the manuscript.

**Disclosures.** The authors declare no conflicts of interest.

**Data availability.** No data were generated or analyzed in the presented research.